\input jytex.tex   
\typesize=10pt
\magnification=1200
\baselineskip17truept
\footnotenumstyle{arabic}
\hsize=6truein\vsize=8.5truein
\sectionnumstyle{blank}
\chapternumstyle{blank}
\chapternum=1
\sectionnum=1
\pagenum=0

\def\begintitle{\pagenumstyle{blank}\parindent=0pt\begin{narrow}[0.4in]}
\def\endtitle{\end{narrow}\newpage\pagenumstyle{arabic}}


\def\beginexercise{\vskip 20truept\parindent=0pt\begin{narrow}[10
truept]}
\def\endexercise{\vskip 10truept\end{narrow}}


\def\eql#1{\eqno\eqnlabel{#1}}
\def\ref{\reference}
\def\peq{\puteqn}
\def\pref{\putref}

\def\mgn{\marginnote}
\def\bex{\begin{exercise}}
\def\eex{\end{exercise}}


\def\mbox#1{{\leavevmode\hbox{#1}}}

\def\hspace#1{{\phantom{\mbox#1}}}

\def\al{\alpha}
\def\be{\beta}

\def\Ga{\Gamma}

\def\om{\omega}

\def\ze{\zeta}

\def\Real{{\rm Re\,}}

\def\zf{$\zeta$--function}
\def\zfs{$\zeta$--functions}


\def\frac#1/#2{\leavevmode\kern.1em
\raise.5ex\hbox{\the\scriptfont0 #1}\kern-.1em/\kern-.15em
\lower.25ex\hbox{\the\scriptfont0 #2}}
\def\sfrac#1/#2{\leavevmode\kern.1em
\raise.5ex\hbox{\the\scriptscriptfont0 #1}\kern-.1em/\kern-.15em
\lower.25ex\hbox{\the\scriptscriptfont0 #2}}

\def\gtorder{\mathrel{\raise.3ex\hbox{$>$}\mkern-14mu
             \lower0.6ex\hbox{$\sim$}}}
\def\ltorder{\mathrel{\raise.3ex\hbox{$<$}\mkern-14mu
             \lower0.6ex\hbox{$\sim$}}}

\def\semidirprod{\rlap{\ss C}\raise1pt\hbox{$\mkern.75mu\times$}}
\def\for{\lower6pt\hbox{$\Big|$}}
\def\fish{\kern-.25em{\phantom{abcde}\over \phantom{abcde}}\kern-.25em}


\def\boxit#1{\vbox{\hrule\hbox{\vrule\kern3pt
        \vbox{\kern3pt#1\kern3pt}\kern3pt\vrule}\hrule}}
\def\dalemb#1#2{{\vbox{\hrule height .#2pt
        \hbox{\vrule width.#2pt height#1pt \kern#1pt
                \vrule width.#2pt}
        \hrule height.#2pt}}}

\def\frac#1#2{{{#1}\over{#2}}}

\def\comb#1#2{{\left(#1\atop#2\right)}}

\def\eg{{\it e.g. }}
\def\ie{{\it i.e. }}

\def\pa{\partial}


  %

\def\sumdasht#1#2{{\mathop{{\sum}'}_{#1}^{#2}}}

\def\3j#1#2#3#4#5#6{\left\lgroup\matrix{#1&#2&#3\cr#4&#5&#6\cr}
\right\rgroup}

\def\man{{\cal M}}

\def\m?{\mgn{?}}

\def\pa{\partial}

\def\beq{\begin{eqnarray}}
\def\eeq{\end{eqnarray}}


\def\cqg#1#2#3{{\it Class. Quant. Grav.} {\bf {#1}} ({#2}) #3}

\def\jmp#1#2#3{{\it J. Math. Phys.} {\bf {#1}} ({#2}) #3}
\def\jpa#1#2#3{{\it J. Phys.} {\bf A{#1}} ({#2}) #3}

\def\np#1#2#3{{\it Nucl. Phys.} {\bf B{#1}} ({#2}) #3}
\def\pl#1#2#3{{\it Phys. Lett.} {\bf {#1}} ({#2}) #3}

\def\prp#1#2#3{{\it Phys. Rep.} {\bf {#1}} ({#2}) #3}
\def\pr#1#2#3{{\it Phys. Rev.} {\bf {#1}} ({#2}) #3}

\def\prD#1#2#3{{\it Phys. Rev.} {\bf D{#1}} ({#2}) #3}

\begin{title}
\vglue 1truein
\vskip15truept
\centertext {\Bigfonts \bf Zero modes, entropy bounds}
\vskip10truept
\centertext{\Bigfonts \bf and partition functions.}
\vskip 20truept
\centertext{J.S.Dowker\footnote{dowker@a35.ph.man.ac.uk}}
\centertext{\it Department of
Theoretical Physics, } \centertext{The University of Manchester,}
\centertext{ Manchester, England} \vskip 20truept
\centertext {Abstract}
\vskip10truept
\begin{narrow}
Some recent finite temperature calculations arising in the investigation of
the Verlinde--Cardy relation are re--analysed. Some remarks are also made
about temperature inversion symmetry.
\end{narrow}
\vskip 5truept
\vskip 60truept
\vfil
\end{title}
\pagenum=0
\section{\bf1. Introduction}
In a recent article Brevik, Milton and Odintsov, [\pref{BMO}], have
evaluated the thermodynamical quantities for quantised scalars, spinors and
vectors on the Einstein universe. The results are compared with the earlier
evaluations of partitions sums by Kutasov and Larsen, [\pref{KandL}] and by
Klemm, Petkou and Siopsis, [\pref{KPS}], one objective being the
investigation of the validity of the Verlinde--Cardy relation. Some
discrepancies were found.

The evaluation of finite temperature field theoretic quantities on the
Einstein universe was actually undertaken earlier, [\pref{DandC, AandD}],
in the
area of quantum field theory on curved spaces and the various high and low
temperature limits determined. The present article draws upon these earlier
investigations to compare with the more recent analyses. The specific point
we wish to look at is the apparent appearance of a term in the internal
energy proportional to the temperature that, [\pref{BMO}], spoils the
Verlinde-Cardy relation.
\section{\bf 2. Thermal scalar on the circle.}

We begin with the low dimensional case of the two-dimensional space--time
$T\times$S$^1$, which is very well investigated. The case of scalars is
quite adequate for our purposes and was considered in [\pref{BMO}]. In some
ways the mechanism will turn out to be generic, which is why it is given here.

In two dimensions, conformal scalars are minimal scalars. The eigenvalues
of $-\nabla$ are $\om_n^2=n^2/a^2$, $(n=0,\pm1,\pm2,\ldots)$ and we note
the existence of a zero mode $n=0$. The point is that, in such a case, it
is not correct simply to ignore the zero mode, when evaluating the internal
energy for example. This would be correct for ordinary multi--particle
statistical mechanics, but in field theory, the existence of a zero mode
implies an infinite tower of states `attached' to each non-zero level. The
effect, [\pref{Dow1}], is to add a term $b_0 kT$ to the internal energy
where $b_0$ is the number of zero modes (here $b_0=1$).

There is no need to go through the general theory, \eg [\pref{DandK, Dow2}],
and the expression for the internal energy is just written down,
  $$
  E=-{1\over12a}+kT+{2\over a}\sum_{n=1}^\infty{n\over e^{n\be/a}-1}\,,
\quad \be=1/kT\,.
  \eql{inte}$$

The first term is the zero temperature (Casimir) value,  the third is the
standard statistical sum over the non-zero modes and the second is the
quantised effect of the zero mode which is the term omitted in [\pref{BMO}],
but not, of course, in Cardy [\pref{Cardy}].

The high temperature form of this expression can be found in various 
standard ways
and one obtains,
  $$
  E\sim {1\over12a}(2\pi aT)^2\,,
  $$
the term linear in $T$ disappears and the objections of [\pref{BMO}] to the
analysis of Klemm {\it et al} [\pref{KPS}], who mention this term and then
argue it away, are no longer valid.

In slightly more detail, in the application of Poisson summation,
  $$
  \sum_{n=1}^\infty f(n)=-{1\over2}f(0) +\int_0^\infty f(t)\,dt 
+2\sum_{m=1}^\infty
  \Real\int_0^\infty f(t) e^{2im\pi t}\,dt\,,
  \eql{poisson}$$
which was also used in [\pref{AandD}], it is the term $-f(0)/2$ that cancels
the $kT$ in (\peq{inte}). This is the basic mechanism.

As a simple observation, which will be useful later, we relate the number
of zero modes to the value of the \zf\ on a $d$--dimensional spatial
section by the standard result,
  $$
  \ze(0)=C_{d/2}-b_0\,,
  \eql{zero}$$
where $C_{d/2}$ is the constant term in the short--time expansion of the
heat-kernel expansion. If $d$ is odd, and space is closed (\eg S$^d$), this
term is zero and the value of $\ze(0)$ is due purely to any zero modes. On the
circle the \zf\ is the Riemann \zf, $\ze(s)=2a^{2s}\,\ze_R(s)$ which 
yields $b_0=1$.

\section{\bf 3. p--forms in higher dimensions.}
On the Einstein universe, T$\times$S$^3$, the only problem is posed by the
gauge invariance of the electromagnetic field but, before giving explicit
expressions, it is better to analyse the extension of photon theory to
higher dimensions in order to get the correct field content.

For a $p$--form on the space-time T$\times\man$, it is easy to show that
that the Obhukov, [\pref{Obhukov}], combination of ghosts reduces to a
simple alternating sum of $p$--forms on $\man$. We may express this
schematically as,
  $$\eqalign{
  \sum_{k=0}^p(-1)^k\,&(k+1)\,\phi_{T\times\man}(p-k)\cr
&=\sum_{k=0}^p(-1)^k\,(k+1)\big(\phi_\man(p-k)\oplus
\phi_\man(p-k-1)\big)\cr
&=\sum_{k=0}^p(-1)^k\,\phi_\man(p-k)\,,}
  $$
where $\phi(p)$ stands for a $p$--form.

It is now possible to use `t\'el\'escopage' to reduce this sum further for
any spectral quantity such as the \zf. If one makes the Hodge decomposition
of the forms on $\man$ into coexact, exact and harmonic types, and then
uses the isomorphism between exact $p$--forms and coexact $(p-1)$--forms,
only the coexact $p$--form quantity survives the summation, together with
the alternating combination of all the harmonic contributions. The coexact
$p$--form on $\man$ describes the physical degrees of freedom of the
system.

The field content is then, schematically,
  $$
  \phi^{\rm tot}(p)=\phi^{\rm CE}(p)\oplus\sum_{k=0}^p(-1)^k\phi^{\rm H}
(p-k)\,,
  \eql{fcont}$$
where CE refers to a coexact form and H to a harmonic one, of which, we
assume, there is only a finite number.

For the spatial section, S$^d$ with $d$ odd, the $p=(d-1)/2$ theory is
conformally invariant when propagated by the de Rham Laplacian and this
constitutes the appropriate generalisation of the photon field.

General and specific information on $p$--forms on spheres exists in
Copeland and Toms [\pref{CandT}] and Dowker and Kirsten [\pref{DandK}] and in
the more mathematical references therein. Interesting comments regarding
heat--kernel expansions are contained in Elizalde {\it et al} [\pref{ELV}].
See also Cappelli and D'Appollonio, [\pref{CandA}].

The only non-zero Betti numbers on the sphere are $\be_0=\be_d=1$ and it is
then straightforward to check, from the \zf\ for the coexact form for
example, \eg [\pref{DandK}], that
there is no constant term in the short-time expansion of the heat-kernel
for the complete field system, the two terms in (\peq{fcont}) cancelling.

It is worthwhile giving a few analytical details. By specialising to the middle
dimension, $p=(d-1)/2$, case ($d$ odd), the coexact \zf, is
  $$
  \ze^{\rm CE}_p(s)=a^{2s}{2\over p!^2}\sum_{n=p+1}^\infty {\big(n^2-p^2\big)
\big(n^2-(p-1)^2\big)   ....(n^2-1)\over n^{2s}}\,,
  \eql{coexzf}$$
$a$ being the sphere radius.

Although we shall not make great use of it, the alternative expression for
this coexact sphere \zf\ derived in [\pref{DandK}] is worth recording here,
  $$
  \ze^{\rm CE}_p(s)=(-1)^p\,\sum_{k=0}^p(-1)^k\,\comb{2k}k\ze_0^{2k+1}(s)\,,
  \eql{dk}$$
where $\ze_0^{k}$ is the \zf\ on the sphere S$^k$ for a scalar field
conformal in $1+k$ dimensions which itself can be represented as the sum of
two Barnes \zfs,
  $$
  \ze_0^k(s)=\ze_B\big(2s;(k+1)/2\mid{\bf 1}_k\big)+
  \ze_B\big(2s;(k-1)/2\mid{\bf 1}_k\big)\,,
  \eql{bar}$$
corresponding to the sum of Dirichlet and Neumann hemisphere quantities,
[\pref{ChandD}].

Another way of writing the coexact \zf\ is, curiously,
  $$
  \ze^{\rm CE}_p(s)={(p+1)^2\over a^2}\,\comb{2p+2}{p+1}\,\ze_0^{2p+3}(s+1)\,,
  \eql{curious}$$
which can be derived from simple manipulation with the degeneracies.
Together with (\peq{bar}) it provides a route to closed, \ie non--summed,
forms for required quantities. However this will not be pursued here.

Returning to (\peq{coexzf}), the summation can be started at $n=1$ and then
the coexact \zf\ expanded in Riemann \zfs\ using, say, Stirling numbers. It
immediately follows that $\ze^{\rm CE}_p(0)=(-1)^{(d+1)/2}$, as needed to
cancel the term from the harmonic piece. This can be seen from equation
(\peq{zero}) with no {\it true} zero modes for the coexact \zf,
(\peq{coexzf}). The harmonic mode contributes an opposite $(-1)^{(d-1)/2}$
to the constant term in the heat-kernel expansion.

The situation is really exactly like that on the circle, S$^1$, discussed
previously. One can incorporate the harmonic piece into the coexact one by
including $n=0$ in the spectrum and splitting the positive and negative
values, $n=0,\pm1,\pm2,\ldots$. The language one would then use is that
there is a zero mode (at $n=0$) which comes in either positively (if $p$ is
even) or negatively (if $p$ is odd).

Since this `zero mode' is a Fock space state it is, presumably, subject to
the same thermalisation as a normal zero mode. This process is actually
built into the construction of the finite temperature \zf\ which is just a
\zf\ on the manifold S$^1\times\man$, [\pref{Dow2,DandKe,Gibbons}],
  $$
  \ze(s,\be)={i\over\be}
  \sumdasht{{m=-\infty\atop \om_n}}{\infty}
{d_n\over(\om^2_n+4\pi^2m^2/\be^2)^s}
  \eql{ftzf}$$
the $\om_n^2$ and $d_n$ being the relevant eigenvalues and degeneracies
on $\man$ and the dash signifying that the denominator should not be zero.
In the present case the specific $d_n$ and $\om_n$ can be read off from the
zero temperature \zf, (\peq{coexzf}) which we shall now write as
  $$
  \ze_\man(s)={1\over p!^2}\sumdasht{n=-\infty}{\infty}
{\big(n^2-p^2\big)\big(n^2-(p-1)^2\big)   ....(n^2-1)\over (n^2/a^2)^s}\,,
  \eql{zem}$$
extended to $p=0$, when the numerator is $1$.

The finite temperature \zf, (\peq{ftzf}), because both summations are now
double sided, can be related to the Epstein \zf\ by differentiation with
respect to $1/a^2$, which is sometimes useful.

One can again appeal to the general theory as given in [\pref{Dow2}] to
write the total internal energy as,
  $$
  E=E_0+(-1)^pkT+\sum_{\om_n\ne0}{d_n\,\om_n\over e^{\be\om_n}-1}\,.
  \eql{pinten}$$
The second term is the effect of the zero mode and $E_0$ is the zero
temperature value,
  $$
  E_0={1\over2}\,\ze_\man(-1/2)\,,
  $$
which is easily evaluated in terms of Bernoulli polynomials, for example.
The very general development given in [\pref{DandK}] could be used, or
equation (\peq{curious}), but I prefer here to work directly and, instead
of finding an expression valid for all $p$, treat the problem as a purely
arithmetic one, $p$ by $p$. Numerical evaluation is easily implemented.

$E_0$ alternates in sign with $p$ and
the values, $-1/12a$ for $p=0$ and $11/120a$ for $p=1$, agree with known
ones. Incidentally, in the limit $p\to \infty$ one finds the following result,
  $$
  E_0(p)\to(-1)^{p+1}\,{1\over\pi^2a}\,,\quad p\to\infty\,.
  $$

Setting $p$ to 1 reproduces the results given explicitly in 
[\pref{DandC,AandD}]
on the Einstein universe and so will not be reproduced here.
\section{\bf 4. High temperature expansions.}

The important high temperature expansions of the various thermodynamic
quantities have been given in [\pref{DandKe,Dow2}] in terms of standard
heat-kernel coefficients for which we can treat (\peq{zem}) as a {\it bone
fide} \zf\ for evaluation purposes. The crucial point, at the moment, is
that the $C_{3/2}$ coefficient is zero, according to our previous
discussion, and so there cannot be a term proportional to $kT$ in the
expansion of the energy. This is {\it confirmed} by looking at the
expression, (\peq{pinten}), for the internal energy directly and applying
Poisson summation, for example.

It is also clear from the alternative representation, (\peq{dk}), that
the zero mode resides only in the S$^1$ part and so the discussion of the
$0$--form  on the two-torus, regarding the effect of this mode, is actually
generic.

There is, however, a price to pay for this `simplification'. Zero modes, on
the spatial section, give rise to a temperature dependent pole in the free
energy, [\pref{Dow2}], and one has to decide what to do about it.
Equivalently, there is a dependence on the scaling length which implies a
contribution to the conformal anomaly. If the quantum field theory is
formally conformally invariant, this will be the only contribution and the
trace condition is the equation of state,
  $$
  PV={1\over d}(E-b_0T)\,,
  $$
which is an equation purely in terms of the {\it non-zero} modes, since the
zero mode contribution to $F$, being volume independent, does not affect
the pressure. However the entropy suffers from the same drawbacks as the
classical Sackur--Tetrode expression, which is a problem this author does
not know how to solve within the present formalism.

Clearly the coefficients of the powers of $kT$ in the expansions can be
readily determined quite generally but this will not be done here. The
expressions (\peq{coexzf}), (\peq{dk}), (\peq{bar}) also show that there
are only a finite number of terms in the high temperature expansions, which
is the same as the old statement that the conformal heat--kernel expansion
terminates on odd spheres. A neat way of seeing this is to relate the
theories on spheres of different dimensions by differentiation, starting
with the flat S$^1$, \eg [\pref{DandC2,Anderson, Camporesi}]. For even
spheres, the heat-kernel expansion does not terminate. This behaviour is
related to the non-existence of a Huyghens principle in odd dimensional
space-times signaled by the appearance of Bessel functions in massless
propagators. To increase the dimension of the sphere by one, a fractional
derivative is needed.

\section{\bf 5. High and low temperature relations.}

Closely related is the connection between the `Planck'  high temperature
expansion and the `Casimir' low temperature form of, say, the total
internal energy. Such a relation was derived and used by Brown and Maclay
[\pref{BandM}] in their elegant treatment of the finite temperature Casimir
effect between plates. The analysis in their paper formed the basis for our
work on the Einstein universe, T$\times$S$^3$, presented in
[\pref{DandC,AandD}] where this relation was also discussed. Some further
remarks and facts were also discussed in Candelas and Dowker [\pref{CandD}]
which was concerned with the relation between vacuum averaged stress-energy
tensors in conformal space-times.

It is clear from the structure of the Epstein \zfs\ that such a temperature
inversion symmetry will hold, in some form, for cavities of rectangular
shape and their periodic counterparts, the tori. A number of calculations
have explicitly verified these rather elementary facts,
[\pref{Wot,RandT,SandT,FandO}]. The symmetry applies for the other flat
space forms as well, [\pref{DandKe,DandB,Unwin1,Unwin2}].

In two dimensions, in statistical contexts, this symmetry is referred to as
modular invariance and corresponds, more or less, to the interchange of the
two cycles on the torus, S$^1\times$S$^1$. It has been very widely analysed,
especially in connection with conformal field theories and an attempt was
made by Cardy [\pref{Cardy}] to extend the notion to higher dimensions by
considering, in particular, free field theories on T$\times$S$^d$.

Because conformal eigenvalues are perfect squares, the relevant quantities,
such as the \zf, have a general torus appearance, apart from the
degeneracies, and therefore there is the possibility of a high--$aT$
low--$aT$ duality which was realised explicitly in
[\pref{DandC,AandD,CandD}] for the three--sphere.

The existence of zero modes is a minor annoyance in this topic and so one
may  restrict most attention to scalar fields. Although we have discussed
the case of the Einstein universe before, it is worth looking at again as a
very specific case.
\section{\bf 6. Thermal on Einstein universe.}

There are a number of starting points for a discussion of temperature
inversion symmetry on the Einstein universe. One way is to take the final
double sum forms for the various thermodynamic quantities, as in
[\pref{BandM,DandC,AandD}], and look at them. Or one can begin with the 
thermal
\zf, as mentioned in [\pref{CandD}],
  $$
  \ze(s,\be)={i\over2\be}\sumdasht{{m=-\infty\atop n=-\infty}}
{\infty}{n^2\over
 \big(n^2/a^2+4\pi^2m^2/\be^2\big)^s}
  \eql{thzf}$$
and force it to be symmetrical. Before proceeding with this, a necessary
technical point has to be brought forward. The free energy is given as the
limit
  $$
  F=-{i\over2}\lim_{s\to0}{\ze(s,\be)\over s}
  \eql{eff}$$
and in the present case, this is finite because (\peq{thzf}) is just a
derivative, with respect to $1/a^2$, of an Epstein \zf\ which equals $-1$
at $s=0$. Hence we don't need to take the limit of (\peq{eff}) any further
in order to analyse the general properties of $F$.

A useful dimensionless variable is $\xi=2\pi a/\be$, and
  $$
 {1\over\xi} aF={1\over4}\lim_{s\to0} {1\over s}\,
 \sumdasht{{m=-\infty\atop n=-\infty}}
{\infty}{n^2\over \big(n^2+\xi^2m^2\big)^s}\,.
  $$
The symmetrisation of this quantity amounts to an evaluation of the internal
energy $E=\pa(\be F)/\pa\be$ or,
  $$
  {1\over\xi^2}\,e(\xi)=-{\pa\over\pa\xi}\bigg({f(\xi)\over\xi}\bigg)\,,
  \eql{diff1}$$
where $e$ and $f$ are defined by the rescaling,
  $$
  f(\xi)=aF\,,\quad e(\xi)=aE\,,
  $$
and we easily see that
  $$
  {1\over\xi^2}\,e(\xi)=\xi^2\,e\big({1\over\xi})\,,
  \eql{tinv}$$
which is the desired temperature inversion of the {\it total} internal energy
and allows high $T$ and low $T$ to be related. To obtain the precise values,
the calculation has to be taken a bit further, [\pref{DandC,AandD}].
It is perhaps fortuitous that the quantity, the energy, that satisfies a
simple temperature inversion relation is a standard thermodynamical one.

One can view the differentiation that occurs in (\peq{diff1}) as equivalent
to the one that raises the dimension of a sphere by two, as was 
mentioned earlier.
That is, the thermal circle has been turned into a `thermal three--sphere'
so that the total manifold becomes symmetrical, effectively S$^3\times$S$^3$,
allowing an inversion symmetry to manifest itself. 

The reason why the relation (\peq{tinv}) is so simple is that the conformal
heat--kernel expansion on the three-sphere terminates with the first,
volume or Weyl, term and the high temperature expansion has only one term
-- the Planck $T^4$ one. Things are slightly different for the higher
spheres. However, before briefly looking at these, it is worth examining
what appears to be another route to temperature inversion symmetry. This
time we take the limit in (\peq{eff}) seriously. Evaluation using heat
kernels, [\pref{DandKe}], or the functional equation  for the Epstein \zf,
[\pref{Kennedy}], can produce the standard statistical mechanical
expression,
  $$
  e(\xi)=e_0+\sum_{n=1}^\infty{n^3\over e^{2\pi n/\xi}-1}\,.
  \eql{sinten}$$
where $e_0=aE_0$ is the zero temperature Casimir constant first evaluated by
Ford, $e_0=1/240=-B_4/8$ ($B_4$ is a Bernoulli number).

One now examines the properties of the sum in (\peq{sinten}) directly.
Early papers in which this was done are by Marlukar [\pref{Marlukar}]
and Glaisher [\pref{Glaisher}]. Let us define the sum under consideration as
  $$
  S(\al,q)=\sum_{n=1}^\infty {n^q\over e^{2\al n}-1}\,.
  $$
Then Marlukar shows anaytically that,
  $$
  S(\al,3)=\bigg({\al\over\be}\bigg)^2\,S(\be,3)+{B_4\over8}\bigg(1-
  \bigg({\al\over\be}\bigg)^2\bigg)\,,
  \eql{marl}$$
where
  $$
  \al\be=\pi^2\,,
  $$
which is precisely the inversion (\peq{tinv}) for the total energy.

There are various ways of organising the information, usually distinguished
by the ways the factors in S$^1\times\man$ are combined analytically.
(S$^1$ here stands for the thermalising imaginary time circle.) One can use
heat-kernels to encode the mode information, or, possibly \zfs. In the
latter case there is a standard and informative way of obtaining the high
temperature expansion of the thermodynamic quantities which is usually
stated for $\log Z$, defined generally by
  $$\ln Z(\be)=-\sum_m\ln\big(1-e^{-\be \om_m}\big)\,, $$
for zero chemical potential. This is to write the Mellin transform,
  $$ \ln Z(\be)={1\over2\pi
 i}\int_{c-i\infty}^{c+i\infty}ds\,\Ga(s)\,
 \ze_\man(s/2)\,\ze_R(1+s)\,\be^{-s}\,, \quad \Real c>d\,, \eql{nrfreen}$$
which arises from the relation between \zfs\ on the product
$\man_1\times\man_2$ and on the factors.

Rather than $\be F_1=-\log Z$, we concentrate on the finite temperature
energy correction, $E_1$,
  $$ E_1={1\over2\pi
 i}\int_{c-i\infty}^{c+i\infty}ds\,\Ga(s+1)\,
 \ze_\man(s/2)\,\ze_R(1+s)\,\be^{-s-1}\,, \quad \Real c>d\,, \eql{nren}$$
which can be looked upon as a continuous expansion in the temperature. The
asymptotic behaviour is determined in an elegant fashion by pushing the
inverse Mellin contour around, usually to the left, and using the
analytical properties of the \zfs. For example, there is always the volume,
or Weyl, pole in $\ze_\man(s)$ at $s=d/2$ giving the Planck term. There is
also a pole at $s=-1$ from the $\Ga$--function which yields the
contribution $-\ze_\man(-1/2)/2$, assuming this quantity is finite,
and this is minus the vacuum Casimir energy, $E_0$, 
[\pref{DandKe,Gibbons,DandB}].

For conformal scalars on the three--sphere, $\ze_\man(s)=a^{2s}\ze_R(2s-2)$
and the high temperature series is rapidly found (as it is in the other
methods). Being exact, (\peq{nrfreen}) can be used to investigate
the the temperature inversion question somewhat explicitly. The quantity of
interest is
  $$
  e_1(\xi)={1\over2\pi
 i}\int_{c-i\infty}^{c+i\infty}ds\,\Ga(s+1)\,
 \ze_R(s-2)\,\ze_R(s+1)\,\bigg({\xi\over2\pi}\bigg)^{s+1}\,,
 \quad \Real c>3\,, \eql{nren2}$$
the poles of the integrand at $s=3$ and $s=-1$ corresponding, as mentioned,
to the Planck and (minus) the Casimir term, $-e_0$. The total energy is
$e(\xi)=e_0+e_1(\xi)$, (\peq{sinten}).

Simple manipulation with the functional equation for the Riemann \zf\
quickly yields the equivalent form,
  $$
  e_1(\xi)=\xi^4{1\over2\pi
 i}\int_{2-c-i\infty}^{2-c+i\infty}ds\,\Ga(s+1)\,
 \ze_R(s-2)\,\ze_R(s+1)\,\bigg({1\over2\pi\xi}\bigg)^{s+1}\,,
 \quad \Real c>3\,, \eql{nren3}$$
and, translating the contours in (\peq{nren2}) and (\peq{nren3}) into 
coincidence,
the inversion symmetry (\peq{tinv}) follows immediately.

This method, which is also employed by Cardy [\pref{Cardy}] and by 
Kutasov and Larsen [\pref{KandL}], is entirely
equivalent to the one involving heat--kernels and theta functions. It has a
certain general disadvantage exemplified by its application to the case when 
$\man$ is a torus. The relevant \zf, $\ze_\man(s)$, is of Epstein form, 
and the standard functional equation is not adequate since in the step to
(\peq{nren3}), the two Riemann \zfs\ are switched. The interchange of
the thermal circle with one of the factors of the torus is brought out most
clearly in the thermal \zf\ method, as in (\peq{thzf}). There {\it is} a
relevant functional relation for the Epstein function based on Rosenhain's
generalisation of the Jacobi transformation for theta functions that allows
one to invert part of the modulus, rather than all of it (which yields the
standard functional relation) but this route seems unnecessarily
complicated.
\section{\bf7. Higher spheres.}

The \zf\ on spheres is an old topic and well investigated. It is clear 
from the form of the standard degeneracies that, in
order to obtain a symmetrical quantity, further differentiations with
respect to $\xi$ must be performed, effectively turning the thermal circle
into a thermal $d$--sphere. Alternatively, the \zf, $\ze_\man$, can be 
expressed as a sum of Riemann \zfs\ and the above analysis performed 
for each piece, producing a sum of terms with different inversion properties.
This piecemeal approach, adopted by Kutasov and Larsen, [\pref{KandL}]
has practical uses.

\begin{ignore}
The standard eigenvalues, $\om_n^2$ are given by
  $$
\om_n=n/a\,,\quad n=2,3,\ldots
  $$
with degeneracy
  $$
  d_n=2(n^2-1)
  $$
and the corresponding \zf\ is (compare [\pref{Dow3}]),
  $$
\ze(s)=2a^{2s}\big(\ze_R(2s-2)-\ze_R(2s)\big)\,.
  $$
Setting $s=0$ confirms the number of zero modes as $b_0=-1$. Note that
$b_0$ is half the degeneracy $d_0$. This is the usual weighting of $0$ in a
symmetrical spectrum labelled from $-\infty $ to $+\infty$. The negative sign
indicates that the zero mode has a ghost-like behaviour.

Hence, according to the general theory, we have for the total internal
energy
  $$
  E={11\over120 a}-kT+{2\over a}\sum_{n=2}^\infty{n(n^2-1)
  \over e^{n\be/a}-1}\,.
  \eql{etotem1}$$

In order to find the high temperature limit, Poisson summation
(\peq{poisson}) is employed. Again the $-f(0)/2$ term cancels the linear $T$
term and leaves equation (41) of [\pref{AandD}],
  $$
   E={2\pi^4a^3T^4\over15}-{aT^2\over3}+{4\over a}\sum_{n=1}^\infty
   \Real\bigg[{\xi^4\over\pi^4}\Ga(4)\ze_R(4,1-2in\xi)-
   {\xi^2\over\pi^2}\ze_R(2,1-2in\xi)\bigg]+{11\over120a}\,.
  \eql{etotem2}$$

By an oversight, the $-kT$ contribution was omitted from the total energy in
[\pref{AandD}], but the form (\peq{etotem2}) is correct and in the 
high $\xi$ limit
the summation tends to, $-11/120a$ \ie minus the zero temperature Casimir term,
so that, up to exponentially small terms,
  $$
  aE\sim {2\pi^4a^4T^4\over15}-{a^2T^2\over3}\,,
  $$
with no qualifications.
\end{ignore}

\section{\bf 8. Conclusion.}

The implications of these results for the validity of the Verlinde--Cardy
formula are unclear as the dependence on the scaling length when a 
zero mode is present appears to
render the value of the entropy subject to some uncertainty. These 
issues will be
discussed in a later communication.
\section{\bf References.}
\begin{putreferences}
\ref{BMO}{Brevik,I., Milton,K.A. and Odintsov, S.D. {\it Entropy bounds in
$R\times S^3$ geometries}. hep-th/0202048.} \ref{KandL}{Kutasov,D. and
Larsen,F. {\it Partition sums and entropy bounds in weakly coupled CFT.}
hep-th/0009244.} \ref{KPS}{Klemm,D., Petkou,A.C. and Siopsis {\it Entropy
bounds, monoticity properties and scaling in CFT's}. hep-th/0101076.}
\ref{DandC}{Dowker,J.S. and Critchley,R. \prD{15}{1976}{1484}.}
\ref{AandD}{Al'taie, M.B. and Dowker, J.S. \prD{18}{1978}{3557}.}
\ref{Dow1}{Dowker,J.S. \prD{37}{1988}{558}.}
\ref{Dow3}{Dowker,J.S. \prD{28}{1983}{3013}.}
\ref{DandK}{Dowker,J.S. and Kennedy,G. \jpa{}{1978}{}.} 
\ref{Dow2}{Dowker,J.S. \cqg{1}{1984}{359}.}
\ref{DandK}{Dowker,J.S. and Kirsten, K.{\it Comm. in Anal. and Geom.
}{\bf7}(1999) 641.}
\ref{DandKe}{Dowker,J.S. and Kennedy,G. \jpa{11}{1978}{895}.}
\ref{Gibbons}{Gibbons,G.W. \pl{60A}{1977}{385}.}
\ref{Cardy}{Cardy,J.L. \np{366}{1991}{403}.}
\ref{ChandD}{Chang,P. and Dowker,J.S. \np{395}{1993}{407}.}
\ref{DandC2}{Dowker,J.S. and Critchley,R. \prD{13}{1976}{224}.}
\ref{Camporesi}{Camporesi,R. \prp{196}{1990}{1}.}
\ref{BandM}{Brown,L.S. and Maclay,G.J. \pr{184}{1969}{1272}.}
\ref{CandD}{Candelas,P. and Dowker,J.S. \prD{19}{1979}{2902}.}
\ref{Unwin1}{Unwin,S.D. Thesis. University of Manchester. 1979.}
\ref{Unwin2}{Unwin,S.D. \jpa{13}{1980}{313}.}
\ref{DandB}{Dowker,J.S. and Banach,R. \jpa{11}{1979}{}.}
\ref{Obhukov}{Obhukov,Yu.N. \pl{109B}{1982}{195}.}
\ref{Kennedy}{Kennedy,G. \prD{23}{1981}{2884}.}
\ref{CandT}{Copeland,E. and Toms,D.J. \np {255}{1985}{201}.}
\ref{ELV}{Elizalde,E., Lygren, M. and Vassilevich, D.V. \jmp {37}{1996}{3105}.}
\ref{Marlukar}{Marlukar,S.L. {\it J.Ind.Math.Soc} {\bf16} (1925/26) 130.}
\ref{Glaisher}{Glaisher,J.W.L. {\it Messenger of Math.} {\bf18} (1889) 1.}
\ref{Anderson}{Anderson,A. \prD{37}{1988}{536}.}
\ref{CandA}{Cappelli,A. and D'Appollonio,\pl{487B}{2000}{87}.}
\ref{Wot}{Wotzasek,C. \jpa{23}{1990}{1627}.}
\ref{RandT}{Ravndal,F. and Tollesen,D. \prD{40}{1989}{4191}.}
\ref{SandT}{Santos,F.C. and Tort,A.C. \pl{482B}{2000}{323}.}
\ref{FandO}{Fukushima,K. and Ohta,K. {\it Physica} {\bf A299} (2001) 455.}
\end{putreferences}
\bye